\documentclass{article} 

\usepackage{graphicx,color}
\usepackage{cite} 

\begin{document}

\title{ Exact solutions for a two-electron
 quantum dot model in a magnetic field and application to more complex sytems}

\author{ M.Taut and H.Eschrig\\
Leibniz Institute for Solid State and Materials Research,
IFW Dresden\\
POB 270116, 01171 Dresden, Germany\\
corresponding author: m.taut@ifw-dresden.de}
\maketitle
\large

\begin{abstract}
We discussed exact solutions of the Schr\"odinger equation 
for a two-dimensional  parabolic confinement potential 
in a homogeneous external magnetic field. 
It turns out that the {\em two}-electron system is exactly solvable 
in the sense, that the problem can be  reduced to numerically solving 
one radial Schr\"odinger equation.
 For a denumerably infinite set of 
values of the  effective oscillator frequency 
$\widetilde{\omega}=\sqrt{\omega_0^2+(\omega_c/2)^2}$  
(where $\omega_0$ is the frequency of the harmonic confinement  
potential and $\omega_c$ is the cyclotron frequency of the magnetic field) 
even analytical solutions can be given.
 Our solutions for 
{\em three} electrons are exact in the strong - and the
 weak correlation limit. 
For quantum dot lattices with Coulomb-correlations between the 
electrons in different dots 
exact solutions are given, provided the lattice constant is large 
compared with the dot diameters. 
We are investigating basic physical properties of these 
solutions like the formation and distortion of Wigner molecules, 
the dependence of the  correlation strength  from 
$\omega_0$ and $\omega_c$, 
and we show that in general  there is no exact Kohn-Sham 
system for the semi-relativistic Current-Density-Functional Theory. \\
\noindent
keywords: quantum dots, exact solutions of Schr\"odinger equation,
Wigner molecules, Current-Density-Functional Theory

\end{abstract}

\maketitle

\newpage
\section{Introduction}
Exact (and somtimes analytical)  solutions of the Schr\"odinger equation
 for realistic few-electron models of the quantum-dot-type
 provide a lot of unique opportunities.
The physical essence of basic physical notions like
the formation and distortion of Wigner molecules (WMs)
and the consequences of inter-dot electron correlations 
in dot lattices 
can be understood more easily than with
numerical brute-force approaches, 
which provide no formulas but only data.
Moreover, these phenomena can be monitored over a wide
range of external parameter values, which allows us to
tune the system continuously between the  weakly and strongly 
correlated regime.
Intermediate and strong correlations are of particular interest,
because for weak correlations there are a lot of mean-field
approaches available.
In particular we can easily see what the basic  difference between
strong correlated systems with low density (Wigner crystal)
and high magnetic fields (fractional quantum Hall systems) is.
Further, exact solutions allow us to check the precision of
approximations like Hartree-Fock and Density Functional approaches
 and to reveal their weak and strong points.
Unlike the comparison of approximate solutions for real systems
with experiments, this approach has the advantage that
all physical quantities
 (including those which are not experimentally accessible)
can be considered, there are no experimental  side effects,
which obscure the comparison, and 
there are no discrepancies due to differences between
 the model and the real system.
Last but not least, basic mathematical assumptions about the structure
of solutions
(like non-interacting $v$-representability in current density functional
theory),
which cannot be proven for general systems,
can sometimes be rejected for special systems.

This paper is {\em not} a comprehensive review 
on  quantum dots and quantum dot molecules and lattices. 
In particular it does {\em not} describe the approaches and the results 
from numerical diagonalizations in a 
complete set of basis functions, 
quasi-classical approaches for the Wigner limit, 
as well as quantum Monte Carlo --, 
density matrix renormalization group --, current density functional --, 
and Hartree-Fock approaches.
Each of these methods warrants a separate review 
(see e.g. Ref.s 
\cite{Hawrylak-98,Maksym-00,Reimann-02,Yannouleas-07}
 and references therein).
Instead, this paper is focused on those systems which can be solved 
exactly or analytically, albeit the more complex sytems only 
in some limits for the external parameters.

A problem with  some exactly solvable models is that
they have to be
 sufficiently simple and the question is whether all their features
are shared by real systems of greater complexity.
Therefore, all approaches complement one another
and they should be pursued in parallel.
Apart from this aspect, our models for the two and three-electron quantum dots
 and lattices from two-electron dots 
are already interesting on its own.

\section{Specification of the model and exact solutions}
\subsection{Model Hamiltonian}
We consider a two-dimensional (2D) two-electron system
(with Coulomb interaction) in
a harmonic scalar potential $v^{ext}(r)=(1/2)\;\omega_0^2 \; r^2$
and a magnetic field ${\bf B}=B \;{\bf e}_z$
represented by the vector potential
(in symmetric gauge)
${\bf A}^{ext}({\bf r})=(1/2) \;{\bf B}\times {\bf r}=
(1/2)\;B\;r\;{\bf e}_{\alpha}$. We introduced cylinder coordinates
$(r,\alpha,z)$ with the cylinder axis perpendicular to the plane, 
to which the electron motion is confined.
The Hamiltonian reads
\begin{eqnarray}
H&=&\sum\limits^2_{i=1}\biggl\{{1\over 2}\biggl({\bf p}_i+
{1\over c}{\bf A}^{ext}({\bf r}_i)\biggl)^2 +
{1\over 2}\;\omega_0^2 \;r_i^2\biggl\}  \nonumber\\
&+&{1 \over |{\bf r}_2-{\bf r}_1|}
+H_{spin}\; ,
\label{H}
\end{eqnarray}
where $H_{spin}=g^*\; \sum\limits_{i=1}^3 \;{\bf s}_i \; \cdot{\bf B}$,
 and atomic units $\hbar=m=e=1$ are used throughout.
This is a widely used effective Hamiltonian
 model for a two-electron quantum dot.

\subsection{Exact solutions of the Schr\"odinger equation}
The Schr\"odinger equation with the Hamiltonian
(\ref{H}) can be solved not only by reduction to the numerical
solution of an (ordinary) radial Schr\"odinger equation \cite{Merkt},
but even analytically for a
 discrete, but infinite set of effective frequencies
$\widetilde{\omega}=\sqrt{\omega_0^2+(\omega_c/2)^2}$ \cite{Taut-2einB},
where we introduced the cyclotron frequency $\omega_c=B/c$.

If we introduce relative and center of mass (c.m.\@) coordinates
\begin {equation}
{\bf r}={\bf r}_2-{\bf r}_1~~~~~,~~~~~{\bf R}={1\over 2}({\bf r}_1+{\bf
r}_2)
\end{equation}
the Hamiltonian (\ref{H}) decouples exactly.
\begin{equation}
H=2 \;H_r+{1\over 2} \;H_R + H_{spin}
\end{equation}
The Hamiltonian for the c.m. motion agrees with the
Hamiltonian of a non-interacting particle in a magnetic field
\begin {equation}
H_R={1\over 2}\biggl[{\bf P}+
{1\over c}{\bf A}_R\biggr]^2+{1\over
2}\omega_R^2\; R^2
\label{H_R}
\end{equation}
and only the relative Hamiltonian contains the
electron-electron interaction
\begin {equation}
H_r={1\over 2}\biggl[{\bf p}+
{1\over c}{\bf A}_r\biggr]^2+{1\over
2}\omega_r^2\;r^2+\frac{1}{2r}\;\; ,
\label{H_r}
\end{equation}
where we introduced rescaled parameters
$\omega_R= 2\omega_0$, ${\bf A}_R=2{\bf A}({\bf R})$,
$\omega_r={1\over 2}\omega_0$, ${\bf A}_r={1\over 2}{\bf A}({\bf r})$
(the indices '$r$'  and '$R$' refer to the relative
and c.m.\@ coordinate systems, respectively).
The decoupling of $H$ allows the ansatz
\begin {equation}
\Phi=\xi({\bf R}) \; \varphi({\bf r})\; \chi(s_1,s_2) \;\; ,
\label{Phi}
\end{equation}
where $\chi(s_1,s_2)$ is the singlet or triplet spin eigen-function.

The eigen-functions of the c.m.\@ Hamiltonian (\ref{H_R})
have the form
\begin{eqnarray}
\xi={e^{iM\cal{A}}\over \sqrt{2\pi}}~~{U_M(R)\over R^{1/2}}=
\frac{e^{iM\cal{A}}}{\sqrt{2\pi}}~~R_M(R) \; ,\nonumber \\
~~~~~M=0,\pm 1,\pm 2,\ldots \;\;,
\label{xi_R}
\end{eqnarray}
where the polar coordinates of the c.m.\@ vector  are denoted by
$(R,\cal{A})$
 and the radial functions $U_M(R)$ and $R_M(R)$
can be found in standard textbooks.

With the following ansatz for the relative motion
\begin{equation}
\varphi={e^{im\alpha}\over \sqrt{2\pi}}~~{u_m(r)\over
r^{1/2}}~~~~~,~~~~~m=0,\pm 1,\pm 2,\ldots
\label{phi_r}
\end{equation}
the Schr\"odinger equation
$H_r\,\varphi({\bf r})=\epsilon_r\,\varphi({\bf r})$
gives rise to a radial Schr\"odinger equation for $u(r)$
\begin{equation}
\biggl\{-{1\over 2}~{d^2\over dr^2}+{1\over 2}\biggl(m^2-{1\over 4}\biggr)
{1\over r^2}+
{1\over 2}\;\widetilde\omega_r^2\; r^2+\frac{1}{2r}
\biggl\}u(r)
=\widetilde \epsilon_r \; u(r)\;\; ,
\label{rad-SGl}
\end{equation}
where the polar coordinates for the relative
 vector are denoted by $(r,\alpha)$,
$\widetilde\omega_r={1\over 2}\widetilde\omega$, 
$\widetilde \epsilon_r=\epsilon_r-{1\over 4}\, m \, \omega_c$, 
and $\omega_c=B/c$.
The solutions are subject
to the normalization condition $\int\limits^\infty_o dr|u(r)|^2=1$.
The Pauli principle demands that (because of
 the different particle exchange symmetry of the spin eigen-functions)
the relative angular momentum
$m$ has to be  even and odd in the singlet and triplet state, respectively.
There is no constraint for the c.m.\@ angular momentum $M$
 following from the Pauli principle.
Because of the orthogonality of the coordinate transformation, the
above described solutions are eigen-functions
 of the total orbital angular momentum
with the eigenvalue $M_L=M+m$.

\begin{figure}[htbp]
\centering
\includegraphics[width=10cm]{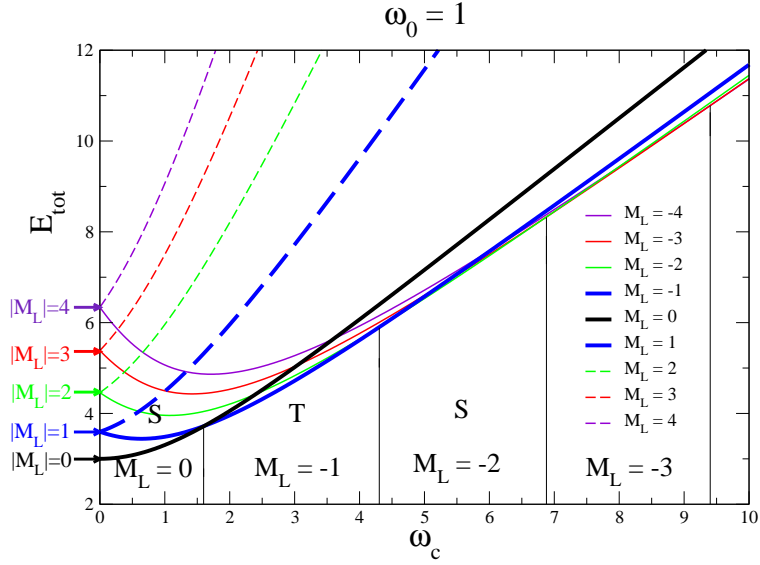}
\caption{
Total energy for fixed confinement frequency $\omega_0=1$
versus cyclotron frequency $\omega_c$ (i.e. magnetic field).
The c.m. system is always in the ground state with $M=0$. The
relative angular momentum $m$ is varied. The vertical lines
show where the
total orbital angular momentum $M_L=M+m$ of the ground state changes.
S and T indicates whether the ground state is singlet or triplet.
Thick lines indicate states which can be NIVR.}
\label{fig-s-t}
\end{figure}

Fig.\ref{fig-s-t} shows that the modulus of the
orbital angular momentum of the ground state (GS) grows stepwise
  with increasing magnetic field.
This implies that the spin state oscillates
between singlet and triplet \cite{Merkt2}.
The Zeeman term and quenching of the singlet state
for higher magnetic fields
 is not included in Fig.\ref{fig-s-t}.
The c.m.\@ excitations are not included as well, 
 because they have no impact on the
character of the ground state.

\begin{figure}[htbp]
\centering
\includegraphics[width=10cm,angle=0]{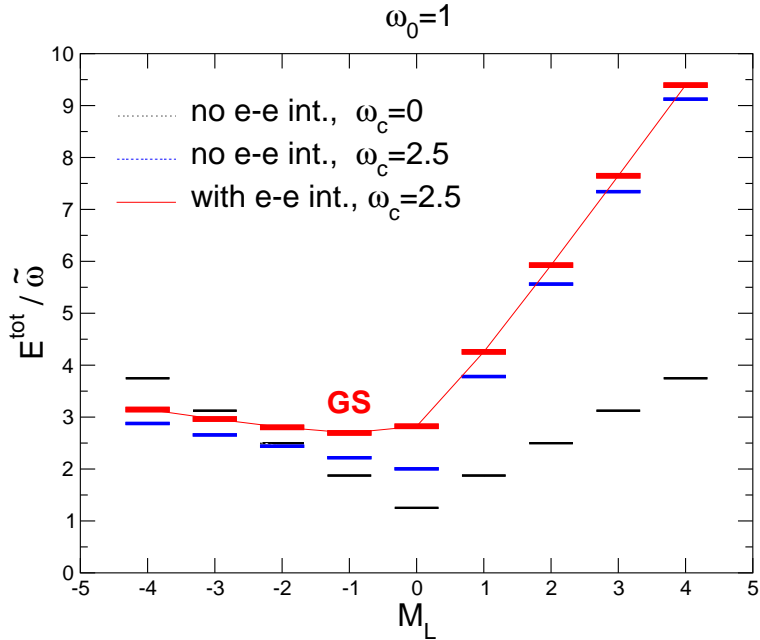}
\caption{
Energy levels for fixed external fields versus total orbital angular momentum.
The level of the GS is indicated.
We started with noninteracting electrons in the confinement only,
added the magnetic field and the e-e-interaction. 
(In the energy unit $\widetilde \omega$ we used 
 $\omega_c=2.5$ for all three cases.)}
\label{fig-E_GS}
\end{figure}

In Fig.\ref{fig-E_GS} the magnetic field and the
e-e-interaction are successively added to the levels in
the confinement only. 
We observe that the magnetic field removes the degeneracy
with respect to the sign of $M_L$  and breaks the symmetry
with respect to up and downward directed fields.
Angular momenta which are parallel to the field (positive)
produce an magnetic moment which is anti-parallel to the field.
They have a large interaction  energy  with the external field
 which shifts the levels upwards. In the opposite case the shift 
due to this contribution is downwards.
Without e-e-interaction the GS has always $M_L=0$.
The shift due to by the
e-e-interaction  is positive definite and it 
decreases with increasing $|M_L|$. 
This can be explained with the radial equation (\ref{rad-SGl}),
which determines the contribution of the relative motion
to the total energy. 
For large $M_L=m$ and small $r$
the last (e-e-interaction) term
is dwarfed by the second (centrifugal potential) term. For large $r$
the third term (effective confinement) is dominating in all cases.
In a classical picture this means that two electrons rotating with a
high angular momentum are separated by the centrifugal force,
so that the job of the e-e-interaction is already largely done and the 
addition of the e-e-interaction does not change much.
It is this $M_L$-dependence of the e-e-interaction shift
 which moves the GS to smaller $M_L$ with
increasing $\omega_c$.

\subsection{Analytical solutions}
\begin{figure}[htbp]
\centering
\includegraphics*[width=12cm,angle=0]{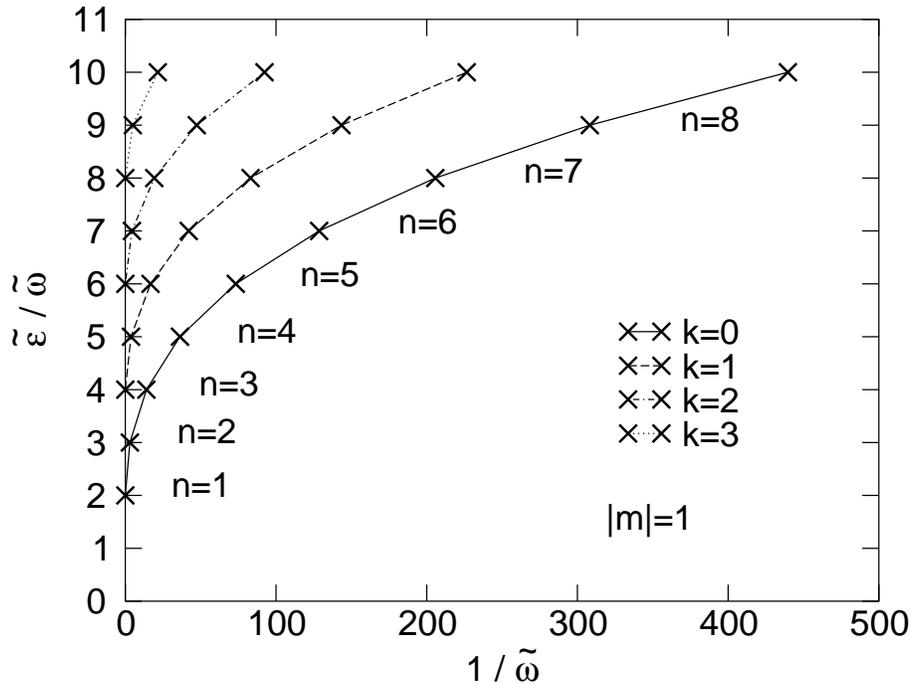}
\caption{
Reduced energies (energy over effective oscillator frequency)
 versus inverse
effective oscillator frequency for  relative angular momentum $|m|=1$.
The crosses indicate solvable states. The lines are just a guide for the eye
and they connect states with the
same node number $k$. $n$ is the same for all horizontal
rows of crosses with the same ordinate.
}
\label{fig-pattern-m=1}
\end{figure}

In \cite{Taut-2einB,Taut-3einB} it has been shown that the
radial Schr\"odinger equation (\ref{rad-SGl})
has simple analytical solutions for a discrete, but infinite set
of effective oscillator frequencies ${\widetilde \omega}$,
the pattern of which for $|m|=1$ can be seen in Fig.\ref{fig-pattern-m=1}.
The patterns for all $|m|$ look qualitatively similar.
All solutions
have the following form:
\begin{equation}
u(r)=r^{|m|+\frac{1}{2}}\; p(r) \; e^{-\frac{1}{4}\; \tilde \omega \; r^2}
\;\;,
\label{u-analyt}
\end{equation}
where $p(r)$ is a {\em finite} polynomial of degree $(n-1)$.
For 'non-soluble' systems, the polynomial has an infinite number of terms.

In this review we
only provide the results for the  simplest analytical solutions.
 For $n=1$ there is only a solution for infinite $\widetilde{\omega}$,
which we call an {\em asymptotic solution}, because it is exact for
${\tilde \omega} \rightarrow \infty$
\begin{equation}
\frac{1}{\tilde \omega}=0 \;\;\; ,\;\;\;
p(r)=1
\label{n=1}
\end{equation}
This solution agrees not only
 with the Laughlin model wave function (WF), if the latter is  applied to $N=2$
and expressed in terms of the coordinates used here,
but it is also the exact solution for non-interacting electrons
(electrons without
Coulomb interaction) in relative- and c.m.\@ coordinates. The
corresponding WF has no node and is a ground state.\\
For $n=2$ there is one {\em finite-field solution}
\begin{equation}
\frac{1}{\tilde\omega}= (2\; |m|+1)  \;\;\; , \;\;\;
p(r) =  1+\frac{r} {(2\; |m|+1)} \;\;,
\label{n=2}
\end{equation}
which is a ground state as well.\\
For $n=3$ there is one asymptotic solution, which is a first excited state,
\begin{equation}
\frac{1}{\tilde \omega}=0 \;\;\; , \;\;\;
p(r)=1-\frac{ r^2}{(|m|+1)}
\label{n=2-excited}
\end{equation}
and one finite field solution, which is a ground state.
\begin{eqnarray}
\lefteqn{ \frac{1}{\tilde\omega}=2\; (4\; |m|+3) \;\;\; ,}
\nonumber\\
&&\;\;\;\;\;\;\;\; p(r) =  1+\frac{r} {(2\; |m|+1)}
          +\frac{r^2}
                {2 (2\; |m|+1)(4\;|m|+3)}
\label{n=3}
\end{eqnarray}
The exact energies $\epsilon$ of the relative coordinate systems
corresponding to these eigen-functions can be obtained
from one compact formula.
\begin{equation}
(\epsilon - \frac{1}{2} m \,\omega_c)=  (|m|+n) \;{\widetilde\omega}
\label{exact-eigenvalues}
\end{equation}
If we compare this result with the spectrum of a single electron
in a quantum dot
 $(\epsilon -\frac{1}{2} m \,\omega_c)= (|m|+2k+1)\; {\widetilde\omega}$,
where $k$ is the node number or degree of excitation,
then it becomes clear that in
both cases $(\epsilon - \frac{1}{2} m \,\omega_c)$ is a integer multiple
of $\widetilde \omega$. We can also say that an analytical solution
exists for those value of $|m|$ and $\widetilde\omega$, for which
one state
of the interacting system is degenerate with one state of 
 the non-interacting system.

A great deal of approaches for
the 2D electron gas in external fields
is based on model wave functions (WF). For inspirations and
checks in the limit N=2 it would be desirable to
have analytical solutions of the two-electron problem
for a wide range of external field values.
One idea is to use the special analytical solutions
(\ref{u-analyt}) with one of  the exactly solvable polynomials given in
(\ref{n=1}-\ref{n=3}) as discussed in Ref.\cite{Taut-3einB}.
If we use a special exact solution for a {\em finite} interval of
external potential values  we
 assume  that the polynomial is independent of $\widetilde \omega$
in this interval
and depends only  on $|m|$.
 Each of the  choices (\ref{n=1}-\ref{n=3})
 provides a different approximation to the exact solution.
Now we check, what the precision of these choices over a wide
range of external fields is.

\begin{figure}[htbp]
\centering
\includegraphics[width=12cm,angle=0]{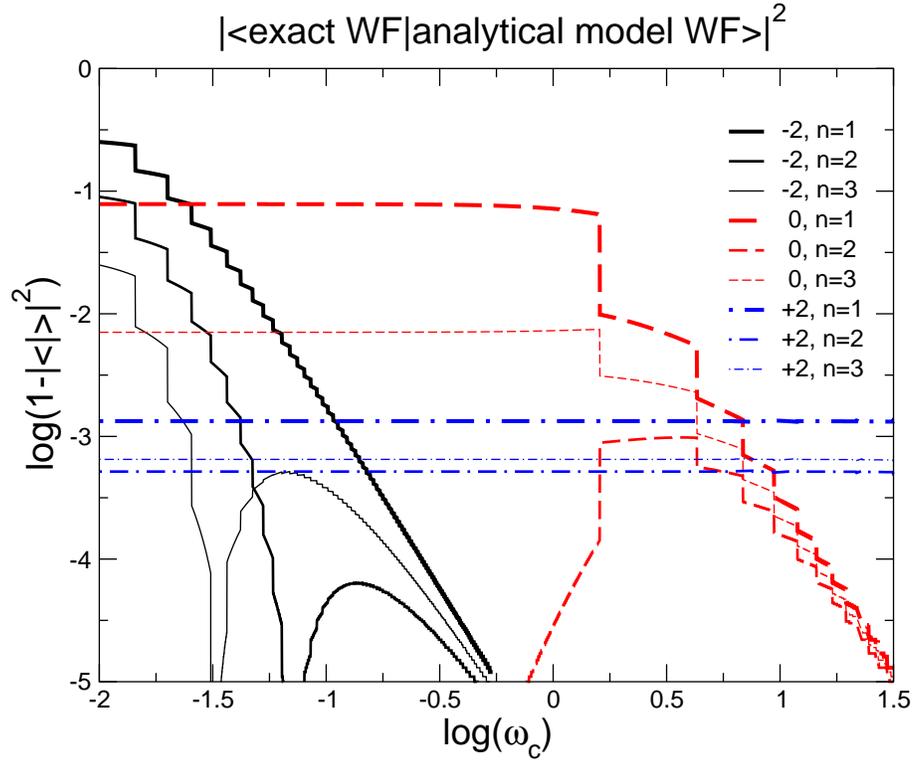}
\caption{
Projection of analytical model wave functions with polynomials
of degree $n$ onto the exact WF.
The logarithm of the deviation of the squared matrix element
from unity is on the ordinate.
Full, dashed  and dash-dotted curves belong to $log(\omega_0)$=--2, 0, and +2,
respectively (see also first number in the legend).
Thick, medium and thin lines belong to n=1,2, and 3,
respectively.
}
\label{fig-pro-lines}
\end{figure}

Fig.\ref{fig-pro-lines}
compares the precision of model WFs with different $n$.
The negative poles in the curves for $log(\omega_0)=+2$ and
n=2 and 3 indicate the vicinity of those $\widetilde \omega$,
which provide exact solutions.
It is seen that the solution for $n>1$ is everywhere better than the solution
for $n=1$
(the latter is exact for infinite fields),
and that all solutions become exact for $\widetilde \omega \rightarrow \infty$.
This means that the ansatz for the N-electron system
 proposed in \cite{Taut-3einB} is definitely more precise than the Laughlin
ansatz, if both are applied to the 2-electron system.
We want to mention that the logarithm of the relative error
 in  the  energies calculated with the
model WFs shows qualitatively the same behavior
as the projection. It is only smaller in magnitude.

\subsection{Exact densities}

With (\ref{xi_R}) and  (\ref{phi_r}), we obtain for the total density
\begin{equation}
n({\bf r})= 2 \int d{\bf r'}\; |\Phi({\bf r},{\bf r'})|^2
\label{def-n}
\end{equation}
the general expression
\begin{equation}
n(r)=\frac{1}{2 \pi^2} \int_0^{2\pi} d\alpha \int_0^\infty dr'\;
\bigg[ R_M\Big(\sqrt{r^2+\frac{1}{4}r'^2+r r' cos\alpha}\Big) \bigg]^2 \;
\Big[u_m(r') \Big]^2
\end{equation}
Because  we are interested in the ground state only, we can safely use
the c.m. state for  $M=0$:
$R_0(R)=2\sqrt{\widetilde{\omega}} \;exp(-\widetilde{\omega} R^2)$
which allows to do one integration analytically leaving us with
\begin{equation}
n(r)=\frac{4 \widetilde{\omega}}{\pi} e^{-2 \widetilde{\omega} \, r^2}
\int_0^\infty dr'\; e^{-(\widetilde{\omega}/2) r'^2} \;
I_0(2 \widetilde{\omega} r r') \Big[u_m(r') \Big]^2
\label{n}
\end{equation}
where $I_n(x)$ are the modified Bessel functions.

The general expression for the paramagnetic current density
\begin{equation}
{\bf j}^p({\bf r})=-i \int d{\bf r'}\;
\Big[
\Phi^*({\bf r},{\bf r'}) \mbox{\boldmath $\nabla$} \Phi({\bf r},{\bf r'}) -
\Phi({\bf r},{\bf r'}) \mbox{\boldmath $\nabla$} \Phi^*({\bf r},{\bf r'})
\Big]
\end{equation}
is somewhat complicated. Therefore, we give here only the formula for
$M=0$
\begin{equation}
{\bf j}^p({\bf r})= {\bf e}_{\alpha}\; m \;\frac{4 \widetilde{\omega}}{\pi}
 e^{-2 \widetilde{\omega} \, r^2}
\int_0^\infty dr'\; e^{-(\widetilde{\omega}/2) r'^2} \;
\frac{I_1(2 \widetilde{\omega} r r')}{r'} \Big[u_m(r') \Big]^2
={\bf e}_{\alpha}\; j^p(r)
\label{j_p}
\end{equation}
As to be expected, the paramagnetic current density is proportional to
the total angular momentum,
 points in azimuthal direction ${\bf e}_{\alpha}$,
and the scalar $j^p(r)$  depends only on the distance $r$ from the center
and not from the azimuthal angle .

Although both formulas (\ref{n}) and (\ref{j_p}) rely on the
functions $u_m(r)$, which are  solutions
of (\ref{rad-SGl}), the analytical behavior for $r \rightarrow 0$
can be expressed in terms of two positive definite integrals.
\begin{eqnarray}
A_0&=&\int_0^\infty dr \; e^{-(\widetilde{\omega}/2) r^2} \;
 \Big[u_m(r) \Big]^2  \\
A_2&=&\int_0^\infty dr \; r^2\; e^{-(\widetilde{\omega}/2) r^2} \;
 \Big[u_m(r) \Big]^2
\end{eqnarray}
After power series expansion of $I_n(x)$, we obtain
\begin{eqnarray}
n(r) &\rightarrow& \frac{4 \,\widetilde{\omega}}{\pi}
e^{-2\, \widetilde{\omega} \, r^2}
\Big[ A_0+A_2 \;\widetilde{\omega}^2 \;r^2 + \cdots \Big] \;\; , \\
\label{n0}\\
j^p(r) &\rightarrow& m \frac{4 \,\widetilde{\omega}^2}{\pi}
e^{-2 \,\widetilde{\omega} \, r^2} \;  r\;
\Big[ A_0+\frac{1}{2} A_2 \;\widetilde{\omega}^2\; r^2 + \cdots \Big]\;\; .
\label{jp0}
\end{eqnarray}
For the origin this means that
$n(0)= 4 \,\widetilde{\omega}\,A_0/\pi$ is always finite and
$j_p(0)=0$  always vanishes.
On the other hand,
the derivative of the density at the origin $\frac{d\, n}{dr}(0)=0$
vanishes, but the derivative of the paramagnetic current density
$\frac{d\, j_p}{dr}(0)=m (4\, \widetilde{\omega}^2 /\pi) A_0$ is finite,
unless $m=0$.
Besides, there is
a relation which does not involve the radial WFs  explicitly.
\begin{equation}
\frac{d \,j_p}{dr}(0)=m\; \widetilde{\omega} \;n(0)
\end{equation}
The exact vorticity, which has the form
$\mbox{\boldmath$\gamma$}({\bf r})={\bf e}_z\;\gamma(r)$ reads
in this limit
\begin{equation}
\gamma(r) \rightarrow m\,2\;\widetilde{\omega} \;\Big(1- \widetilde{\omega}^2 \;
\frac{A_2}{A_0} \; r^2 +\cdots \Big) \;\; .
\label{gamma0}
\end{equation}
As will be seen in Sect.4, the limit $r\rightarrow0$ is decisive for our proof 
of the violation of non-interacting $v$ representability.

\section{Formation of Wigner molecules and correlation strength}
A illustrative classical picture for a WM 
 in an environment with rotational symmetry
is a rotating and vibrating electron molecule.
For a two-electron system this is a dumbbell-like object.
We are going to show that this configuration is a 
manifestation of strong e-e-correlations and it is formed 
in the limit of {\em small}
$\omega_0$ or {\em large} $\omega_c$.
The issue is: why is {\em small}
$\omega_0$ equivalent to {\em large} $\omega_c$ although 
 the exact WF and consequently all 
distribution functions depend {\em only} on the effective confinement frequency 
$\widetilde{\omega}=\sqrt{\omega_0^2+(\omega_c/2)^2}$ 
where both $\omega_0$ and $\omega_c$ 
have qualitatively the {\em same} influence.
In particular, we will point out, how
strong magnetic fields can  cause strong correlations. This is not
the same mechanism as for weak confinement (see also Ref.\cite{Matulis}).

\begin{figure}[htbp]
\includegraphics[width=10cm,angle=270]{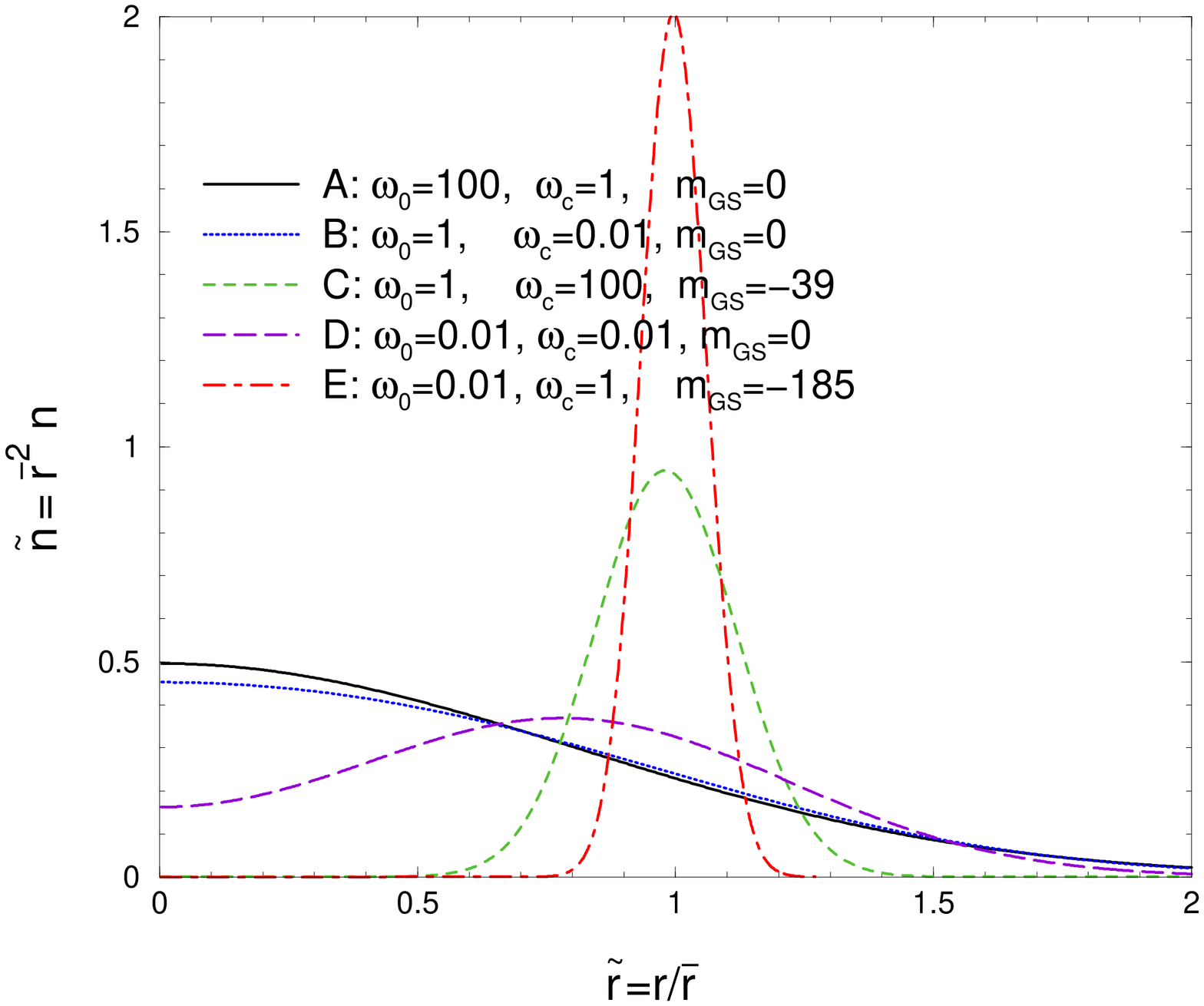}
\includegraphics[width=10cm,angle=270]{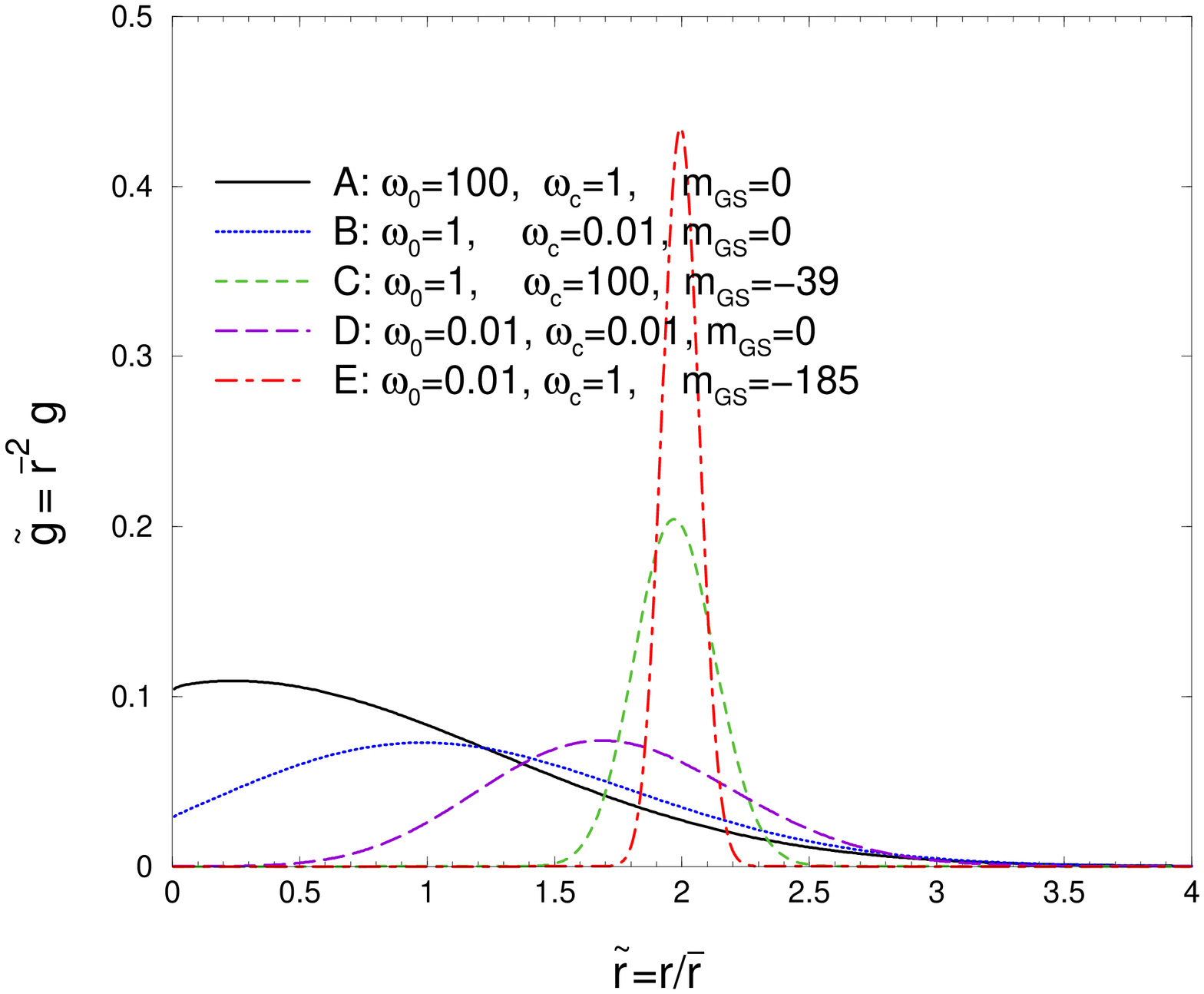}
\caption{ Density (upper)
and pair correlation function (lower)
for several correlation strength. $\omega_0$, $\omega_c$, and the
angular momentum of the ground state are given in the legend.
The scaling parameters are:
$\bar{r}$=0.0901(A), 1.0192(B), 0.6345(C), 14.5022(D), and 13.6472(E)}
\label{fig-n-g}
\end{figure}

For illustration we use the density (\ref{def-n}), which
provides the distribution of the electrons in space,
 and the  pair correlation function
\begin{equation}
 g({\bf r})=<\psi|\sum\limits_{i<j}\delta({\bf r}_i-{\bf r}_j-{\bf r})|\psi>
\label{def-G(r)}
\end{equation}
which determines the distribution of the distance  between two electrons.
Both quantities depend only from the modulus: 
$n({\bf r})=n(r)$ and $g({\bf r})=g(r)$.
In our system the latter is  given by
the radial part of the relative WF alone.
\begin{equation}
g(r)=|\varphi({\bf r})|^2={[u(r)]^2\over 2\pi r}
\label{G(r)}
\end{equation}
In Fig.\ref{fig-n-g} both quantities are shown
for a few typical parameter sets.
We have chosen scaled axes (tilted quantities)
 which allow us to show and compare
different extreme cases in the same picture.
On the abscissas the $r$-coordinate is scaled $\widetilde{r}=r/\bar{r}$
with the average radius of the dot
$\bar{r}=(1/2) \int d^2{\bf r}\;r\; n(r)$ 
and the ordinates are scaled
in such a way that the norms are conserved:
 $\int d^2\widetilde{\bf r} \; \widetilde{n}(\widetilde{r})=2$ and
$\int d^2\widetilde{\bf r} \; \widetilde{g}(\widetilde{r})=1$.
For weak correlations $n(r)$ is peaked
 in the region of lowest potential energy
(center of the dot) and $g(r)$ is spread out over the whole
range of non-vanishing density, allowing in particular
small distances between the electrons.
This regime is realized in  curves A and B
applying to strong ($\omega_0$ large) or medium confinement and
small and medium  $\omega_c$. Strong correlations are connected with
sharply peaked densities at non-zero $r$ confining the electrons on
a ring with radius $\bar{r}$. The pair correlation function is
sharply peaked at a distance $2 \bar{r}$ which is the diameter of the ring.
This means that the electrons are localized at a ring and have virtually
antipodal positions \cite[Taut-2einB] what agrees with the 
above mentioned calssical picture of a WM. 
This minimizes the e-e-interaction energy 
in the limits allowed by the potential confinement energy without 
enhancing the kinetic energy.
In terms of external parameters this can  be realized in two scenarios.\\
i) If  $\omega_c$ is small (or zero),
the confinement has to be weak ($\omega_0$ small).
This can be concluded from the comparison of curve B and D.
For weak confinement the state is spread out widely and the density is low.
Low density implies the  dominance of the the e-e-interaction over
 the kinetic energy
(last term in radial Schr\"odinger equation (\ref{rad-SGl})
versus the first term)
and a state which minimizes the e-e-interaction.
The condensed matter analog to this state is the Wigner crystal.
All in all, {\em  strong correlations in systems with low densities
are produced by the  dominance of the the e-e-interaction over
 the kinetic energy.} \\
ii)  If  $\omega_c$ is medium or large (and the confinement not too strong),
 then the angular momentum of the GS is strongly negative
(see  curve C and E).
Large (modulus of) angular momentum means strong centrifugal potential
(second term in radial Schr\"odinger equation (\ref{rad-SGl}))  which
drives the electrons away from the center and
 produces the ring structure in the density.
At the same time the remaining e-e-term maximizes the e-e-distance within the
limits set by the density. This is most pronounced in curves C and E.
The case of small
$\omega_0$ and large $\omega_c$ (which is not shown because of
its numerical difficulties  caused by of the extremely strong angular momentum)
 is even  more strongly correlated.
The condensed matter analog to this state is the fractional quantum Hall state.
The state of curve C shows strong correlations as well,
but does not have a large diameter
($\bar{r}=0.6345$) and consequently a low density.
This proves that low density is not necessary for strong correlation, but
a magnetic field can do the job alone. All in all,
{\em strong correlations in high magnetic fields are mainly
produced by the high modulus of the angular momentum of the
ground state.}

\begin{figure}[htbp]
\includegraphics[width=14cm,angle=0]{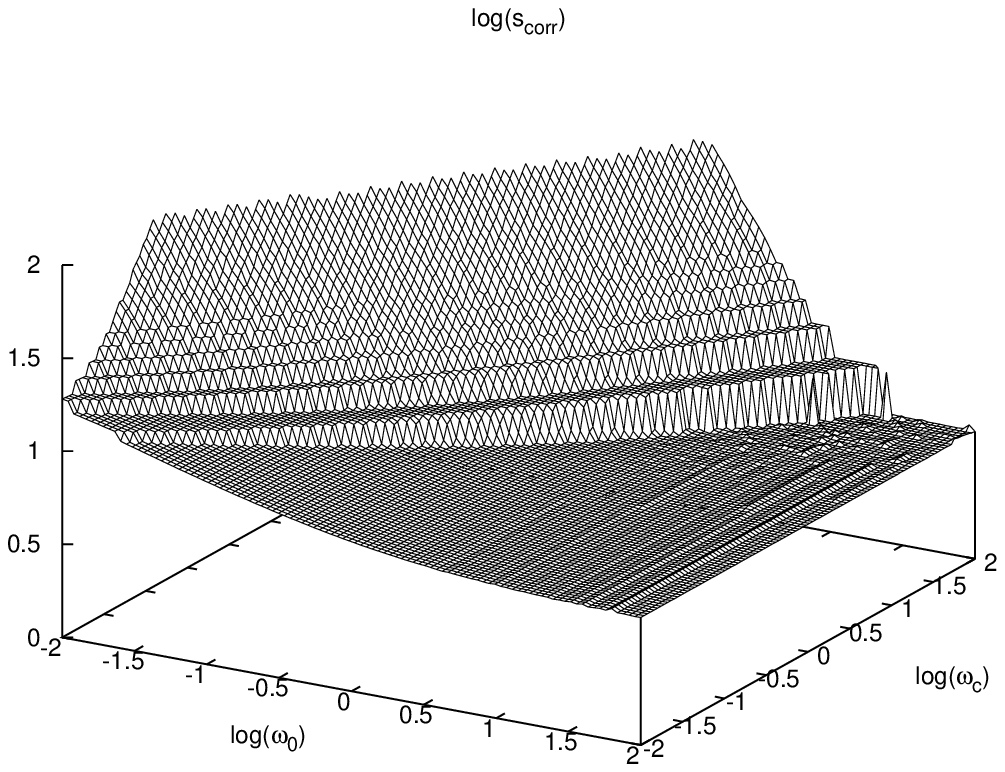}
\includegraphics[width=12cm,angle=0]{corr-lines.eps}
\caption{Correlation strength versus external field parameters in
log-scales for all axes.
The steps, corresponding to parameters where the angular momentum of the GS
changes (see Fig.\ref{fig-s-t}), 
are fully resolved only in the lower panel, which shows curves for 3 discrete 
$\omega_0$.
}
\label{fig-s-corr}
\end{figure}

From these considerations it follows that
in our family of systems a suitable dimensionless
quantitative definition of the 
correlation strength\footnote{Unlike in common quantum chemistry language, 
the {\em correlation strength} defined here comprises
all effects beyond the Hartree approximation, in particular it includes 
the effect of exchange.}
 can be set up by the
mean square radius of the pair correlation function
 $\overline{r^2}=\int d^2{\bf r}\,r^2\, g(r) $ and
its half width
 $\Delta^2 r=\int d^2{\bf r}\, (r-\bar{r})^2 g(r) $, 
where  $\overline{r}=\int d^2{\bf r}\,r\, g(r) $, according to 
\begin{equation}
s_{corr}=\frac{\overline{r^2}}{\Delta^2 r}\;\; .
\label{c-corr}
\end{equation}
This means, small half width and a peak at large $r$ 
produce strong correlations.
Fig.\ref{fig-s-corr} shows this quantity as a function of the
external field parameters.
The steps are caused by a change in the angular momentum of the GS.
It is obvious that for small $\omega_c$ a weakening of the confinement
 increases $s_{corr}$, but an increase of
the magnetic field is much more effective,
{\em if} it is connected to an increase of the modulus of
the angular momentum of the GS, what happens in the region 
where the steps are found.

Matulis and Peeters \cite{Matulis} investigated  the same issue 
using qualitative asymptodic expressions for the wave function 
instead of exact solutions. Nevertheless the resulting trends are the same, 
although their visual picture is different.

\section{Violation of non-interacting $v$-representability of the
exact solutions of the Schr\"odinger equation}

The exact solutions of the special system considered in this review
 can be used to show that (unlike generally assumed)
 an exact Kohn-Sham (KS) system in the framework of
semi--relativistic Current Density Functional Theory (CDFT)
can exist only in special cases.

In  Density Functional Theory (DFT) it can be shown that  for the GS
the external potential is a functional of the density \cite{HK}
(see also textbooks  \cite{Eschrig,Dreizler-Gross}
with more modern approaches)
\begin{equation}
v^{ext}({\bf r}) \;\;
 \stackrel{\cal C}{\leftarrow}
\; \Psi \;\;
\begin{array}{c}
{\scriptstyle \cal D} \\[-2mm]
\rightleftharpoons\\[-2mm]
{\scriptstyle {\cal D}^{-1}}
\end{array}
\;\;n({\bf r})
\label{map}
\end{equation}
which would imply  non-interacting $v$-representability (NIRV)
or the existence of an exact Kohn-Sham system for the GSs, 
if the interacting and the non-interacting systems would 
have a common set of ground state densities.

In the presence of a magnetic field and for (semi-relativistic)
 Current Density Functional Theory (CDFT),
the generalization of ${\cal D}^{-1}$ for the ground state
still exists, but
Vignale and Rasolt \cite{Vignale-Rasolt1, Vignale-Rasolt2}
just {\em presupposed} the existence of the generalization of $\cal C$
 \cite{Capelle-Vignale1} implying that NIVR and the existence
of a KS scheme  has not been proven.
Capelle and Vignale \cite{Capelle-Vignale1}, on the other hand,
have shown that there can be several external potentials ${\cal V}^{ext}$
which provide the same WFs and densities
\begin{equation}
\begin{array}{c}{\cal V}^{ext}_1({\bf r})\\{\cal V}^{ext}_2({\bf r})\\ \cdots \end{array} \;\;
\begin{array}{c} \searrow\\ \rightarrow\\ \nearrow \end{array}\;\;
{\bf \Psi}
\rightleftharpoons
{\cal N}({\bf r})
\label{CDFT-maps}
\end{equation}
where ${\cal V}^{ext}({\bf r})$ and ${\cal N}({\bf r})$ represent
 both external potentials ($v^{ext}({\bf r})$
and $\bf A^{ext}({\bf r})$)
and both densities ($n({\bf r})$ and ${\bf j}^p({\bf r})$), respectively.
Hence, $\cal C$ cannot exist anymore as an unique mapping and the question
of NIVR cannot be answered in this way.
However,
the exact solutions of the special system considered in this review
 can be used to show that
 an exact Kohn-Sham system  or NIVR can exist only in the 
following special cases.

All those states at non-zero $B$ can be  NIVR, which are 
continuously connected  to the singlet and triplet ground  states 
 at $B=0$ (see also Fig.\ref{fig-s-t}). In more detail:

If the GS is a {\em singlet} (total orbital angular momentum $M_L$ is even)
both densities can be  NIVR if the vorticity
 $\mbox{\boldmath$\gamma$}({\bf r})=\mbox{\boldmath $\nabla$}
\times \Big ({\bf j}^p({\bf r}) / n({\bf r}) \Big) $
of the exact solution vanishes.
For $M_L=0$ this is trivially guaranteed
 because the paramagnetic current density vanishes.
The vorticity based on the exact solutions
for the higher $M_L$ does not vanish, in particular for small $r$.
In the limit $r\rightarrow 0$ this can even be shown analytically.

If the GS is a {\em triplet} ($M_L$ is odd) and we
assume circular symmetry for the KS system
(the same symmetry as the real system)
then only the exact states with $|M_L|= 1$ can be NIVR
 with KS states having angular momenta $m_1=0$ and $|m_2|=1$.\\
Without specification of the symmetry of the KS system
the condition for NIVR is that the small-$r$-exponents of the KS states
are 0 and 1.

\begin{figure}[htbp]
\centering
\includegraphics[width=12cm]{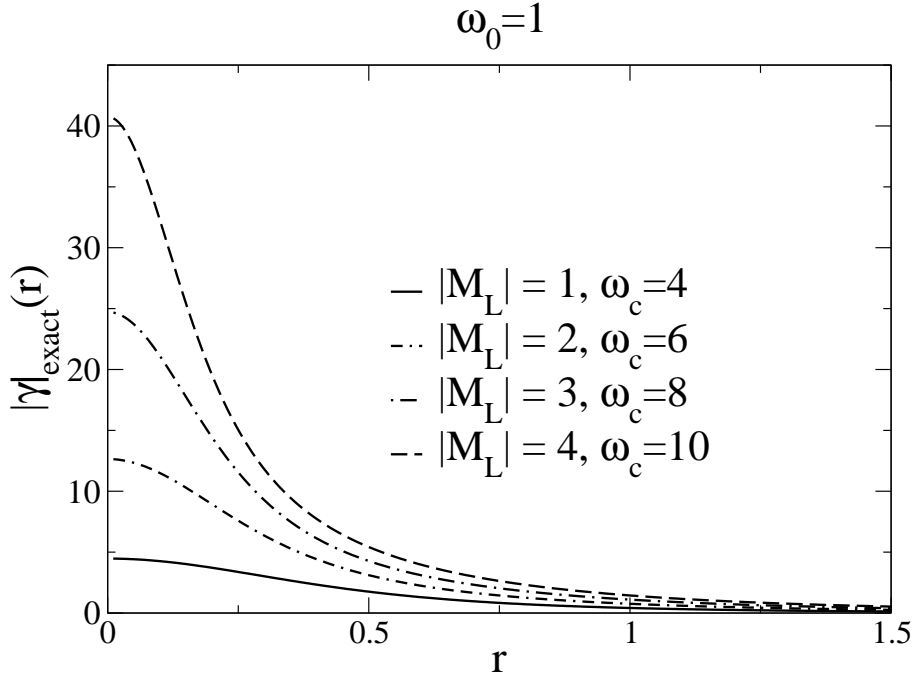}
\caption { The vorticities for $\omega_0=1$ and a few
typical cyclotron frequencies $\omega_c$
 where the state with negative $M_L$ is the ground state.
The sign of $\gamma_{exact}(r)$ agrees with the sign of $M_L$.}
\label{fig-vort}
\end{figure}

The proof of the statement for the singlet state is extremely simple
and will be given here.
The other proofs and more detailed information can be found in \cite{Taut-CDFT}.
The question is if the (in this case doubly occupied)
KS-WF $\varphi({\bf r})=R({\bf r}) \; e^{i\zeta({\bf r})}$ can be chosen
in such a way that the density and the gauge invariant vorticity
of the non-interacting KS system and the exact solution agree.
\begin{equation}
n_{exact}(r) \stackrel{!}{=} n_{KS}({\bf r})=2 \;[R({\bf r})]^2
\label{det-n}
\end{equation}
\begin{equation}
\mbox{\boldmath$\gamma$}_{exact}({\bf r})\stackrel{!}{=}
\mbox{\boldmath$\gamma$}_{KS}({\bf r})=0
\label{det-vort}
\end{equation}
Eq. (\ref{det-n}) defines the real part of the KS-WF.
On the other hand, the vorticity of a two-electron singlet KS state
vanishes exactly irrespective of the special form of
 $R({\bf r})$ and $\zeta({\bf r})$.
Therefore,
equation (\ref{det-vort}) can only  be satisfied if the
vorticity of the corresponding exact solution vanishes as well.
Fig.\ref{fig-vort} shows that this is not the case, in particular for small
$r$ the violation is massive. 
Eq. (\ref{gamma0}) provides $\gamma_{exact}(0)=2\; M_L\; {\widetilde \omega}$ 
which shows that the 'degree of violation' grows with growing 
$\widetilde \omega$.
The exact vorticity vanishes only for the state with zero angular
momentum, which is the GS for small magnetic fields
(see also Fig.\ref{fig-s-t}).

\section{Distortion of the three-electron \\Wigner molecule}
\begin{figure}[]
\begin{minipage}{6.0cm}
\begin{center}
\includegraphics[width=5.7cm,angle=0]{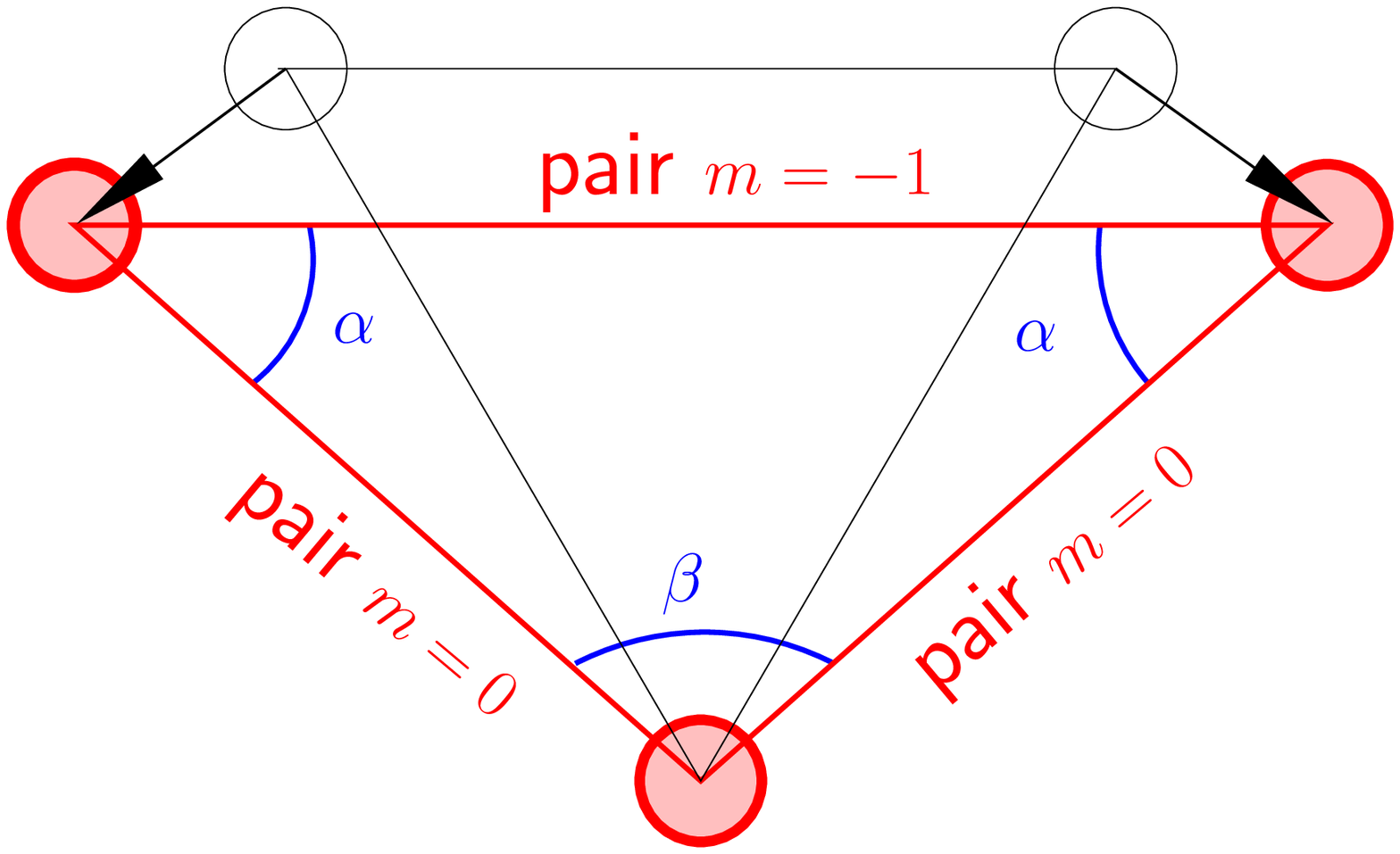}
\end{center}
\end{minipage}
\begin{minipage}{6.0cm}
\begin{center}
\includegraphics[width=4.cm,angle=0]{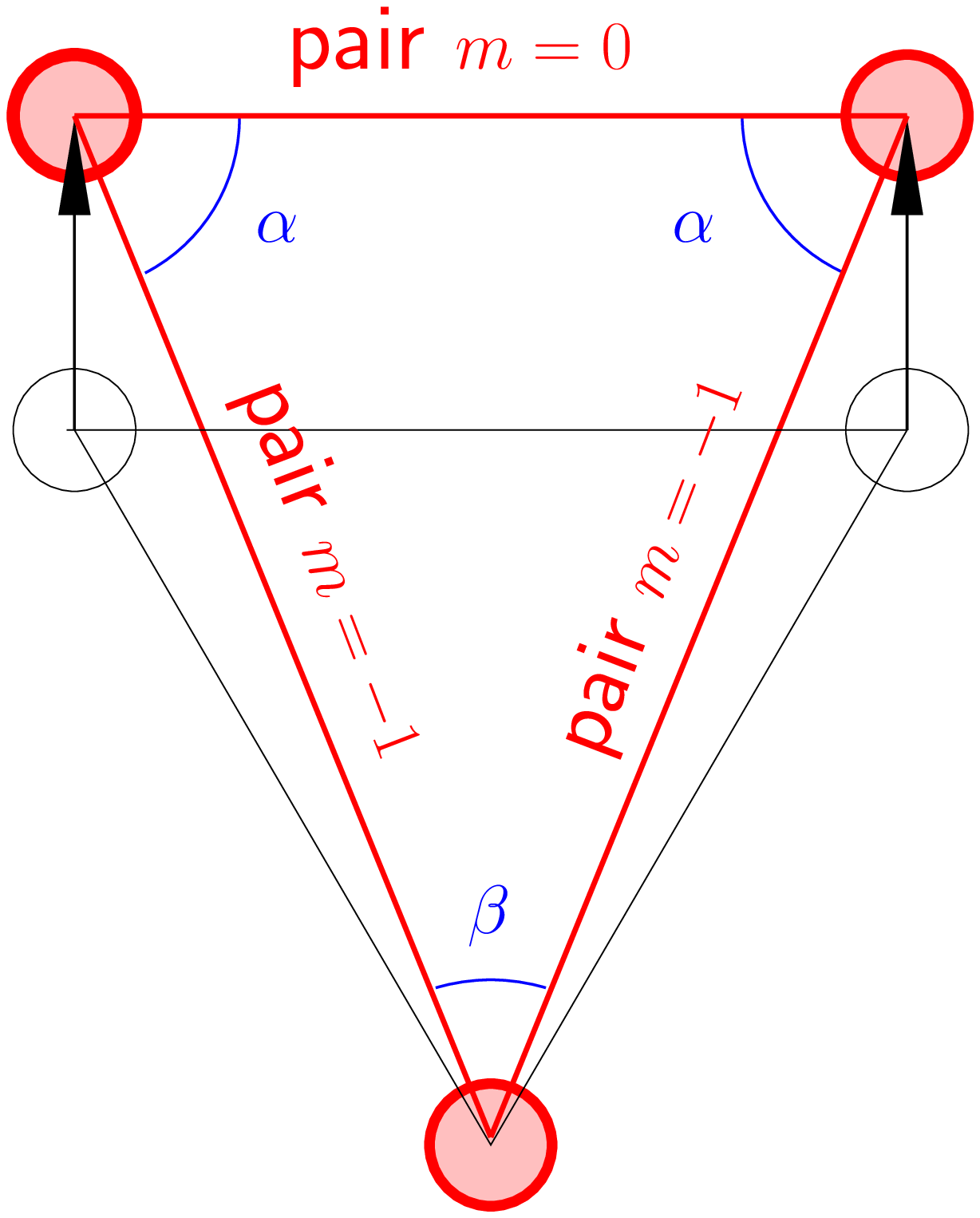}
\end{center}
\end{minipage}
\caption[]{
 Schematic picture  of  the
distorted three-electron Wigner molecule
with total orbital angular momentum $M_L=-1$ (left) and $M_L=-2$ (right).
The thin lines depict the undistorted WM.
}
\label{fig-scetch}
\end{figure}

 As shown above, the two-electron system is the simplest system which
exhibits the phenomenon of the formation of WMs in finite systems.
For three electrons there is another effect, 
namely a Jahn-Teller-like distortion
of the WM \cite{Taut-JahnTeller}
 shown schematically in Fig.\ref{fig-scetch},
which can be investigated using the solutions described above.
The point is that in the strong correlation limit  
the three-electron system can be decoupled into 
three independent pairs, the Schr\"odinger equation for which 
agrees 
(apart from a renormalisation of the interaction parameters)
with the Schr\"odinger equation for the relative coordinates in 
the two electron system \cite{Taut-3einB, Taut-JahnTeller}.

The Hamiltonian of the three-electron system reads
\begin{equation}
H=\sum_{i=1}^3\biggl[{1\over 2}\biggl(\frac{1}{i}{\bf \nabla}_i
+{1\over c}{\bf
A}({\bf r}_i)\biggr)^2 +{1\over 2}\omega_o^2\;r_i^2\biggr]
+\sum_{i<k}{1 \over |{\bf r}_i-{\bf r}_k|}
\label{h-orig}  
\end{equation}
where the the Zeeman term is disregarded because it has no influence on 
the spacial distribution of the electrons, but shifts only the energies.
We consider the unitary coordinate transformation
from the original position
vectors ${\bf r}_i$ to new ones ${\bf x}_i$
\begin{equation}
 \left[ \begin{array} {c} {\bf x}_1\\{\bf x}_2\\{\bf x}_3\end{array} \right]=
\left[ \begin{array} {ccc}1/3&a&b\\b& 1/3&a\\a&b&1/3\end{array}\right]
\left[ \begin{array} {c} {\bf r}_1 \\ {\bf r}_2 \\ {\bf r}_3 \end{array} \right]
\label{trafo} 
\end{equation}
where $a=1/3-1/\sqrt{3}$ and $b=1/3+1/\sqrt{3}$. 
The corresponding inverse transformation provides for the
difference coordinates in the e-e-interaction terms 
\begin{equation}
{\bf r}_i-{\bf r}_j=\sqrt{3}\; \biggl({\bf X}-{\bf x}_k\biggr)
\label{diff} 
\end{equation}
where $(i,j,k)=(1,2,3)$ and cyclic permutations,
and  ${\bf X} \equiv \frac{1}{3} \sum_{i=1}^3 {\bf x}_i$
is the center of mass (c.m.) in the new coordinates.
It is a special feature of this transformation that the latter
agrees with the c.m. {\bf R} in the original coordinates.
This transformation provides the equivalent Hamiltonian 
\begin{equation}
H=\sum_{i=1}^3\biggl[{1\over 2}\biggl(\frac{1}{i}{\bf \nabla}_i+{1\over c}{\bf
A}({\bf x}_i)\biggr)^2 +{1\over 2} \; \omega_o^2 \; x_i^2
+\frac{1}{\sqrt{3}} \; {1 \over |{\bf x}_i-{\bf X}|} \biggr]
\label{h-trans} 
\end{equation}
Next we have to  observe that in the strong correlation limit  
the uncertainty of the c.m. $\bf X$ is small
compared with the expectation value of the new coordinates ${\bf x}_i$
 (see the appendix of Ref.\cite{Taut-JahnTeller}).
Therefore $\bf X$ can be considered as a small perturbation 
and in zero order in  $\bf X$ the Hamiltonian (\ref{h-trans}) decouples 
into a sum of independent quasi-particle Hamiltonians
\begin{equation}
H^{(0)}=\sum_{i=1}^3 h_i
\label{H0}  
\end{equation}
The Schr\"odinger equation for the quasi-particles
\begin{equation}
h\; \varphi_{q}({\bf x}) = \varepsilon_{q} \; \varphi_{q}({\bf x})
\label{pair-eq} 
\end{equation}
is similar to the Schr\"odinger equation for the Hamiltonian 
in the  relative coordinates (\ref{H_r}) and therefore it 
can be solved exactly. 
In terms of these solutions, 
the total energy is a sum of quasi-particle energies and the
total orbital eigenfunction
is a product of quasi-particle functions. 
\begin{eqnarray}
E_{q_1,q_2,q_3}&=&\varepsilon_{q_1}+\varepsilon_{q_2}+\varepsilon_{q_3}
\label{E-tot}\\  
\Phi_{q_1,q_2,q_3}({\bf x}_1,{\bf x}_2,{\bf x}_3)&=&\varphi_{q_1}({\bf x}_1)
\cdot \varphi_{q_2}({\bf x}_2)\cdot \varphi_{q_3}({\bf x}_3)
\label{WF-tot} 
\end{eqnarray}
where $q_i$ comprises all quantum numbers. 
We can consider the quasi-particles as electron pairs,
wherby their WFs $\varphi_{q}({\bf x}_k) $
describe the distance ${\bf x}_k$ between two electrons (see Eq. (\ref{diff})).
The crucial point in explaining the distortion of the WM 
is the generalised Pauli exclusion principle for the 
quasi-particles states, 
i.e., the rules which determine the allowed combination of quantum numbers
in Eqs. (\ref{E-tot}) and (\ref{WF-tot}) in order to 
guarantee the anti-symmetry of the WF under electron transposition.
These rules depend on the total spin ($S$) and orbital angular momentum ($M_L$)
 and they rule for some configurations the agreement of 
all three quantum numbers $q_i$ out 
(see \cite{Taut-3einB}). This means that the 
 electron distances in all three electron pairs 
cannot agree giving rise to a distortion of the WM. 

The final result is the following \cite{Taut-JahnTeller}:\\
In the ground state the electrons in an WM 
form an equilateral triangle (as might be expected from naive reasoning) only,
if the state is a  quartet ($S=3/2$) and the orbital angular momentum
is a magic quantum number ($M_L=3\; m ; m=$ integer). Otherwise
the triangle in the ground state is isosceles.
For $M_L=(3 m+1)$ one of the sides is
longer and for $M_L=(3 m-1)$
one of the sides is shorter than the other two.

\section{ Coulomb correlations between Quantum dots }
Other systems where the above described 
two-electron solutions play a crucial role 
are two-electron  quantum dot molecules and quantum dot lattices, 
where the Coulomb correlation 
between electrons in differents dots is taken into account 
in the Van der Waals approximation \cite{Taut-QD-lattice}.
This means  that the diameter of the dots must be small compared with the 
distance between the dots and that the overlap between  WFs of different 
dots should be  negligible. 
Then the Coulomb interaction between the electrons
at ${\bf r}_{nk}={\bf R}_n^0 + {\bf u}_{nk}$ and 
${\bf r}_{n'k'}={\bf R}_{n'}^0 + {\bf u}_{n'k'}$
in {\em different} dots centered at ${\bf R}_n^0$ 
and ${\bf R}_{n'}^0$ with ($n\neq n'$)
 can be expanded in second order (dipole approximation) as 
\begin{eqnarray}
&&\frac{1}{|{\bf r}_{nk}-{\bf r}_{n'k'}|}=
\frac{1}{|({\bf R}_n^0-{\bf R}_{n'}^0)
    +({\bf u}_{nk}-{\bf u}_{n'k'})|} = \nonumber \\
&&= \frac{1}{|{\bf R}_n^0-{\bf R}_{n'}^0|} +
\frac{1}{2} ({\bf u}_{nk}-{\bf u}_{n'k'})
 \cdot {\bf T}({\bf R}_n^0-{\bf R}_{n'}^0) \cdot
({\bf u}_{nk}-{\bf u}_{n'k'}) +... \nonumber
\end{eqnarray}
wmiltoinanshere the dipole tensor
$ {\bf T}({\bf R})=(1/R^5) [\;3\; {\bf R} \circ {\bf R }
- R^2 \;{\bf I} \;] $
has been introduced. 

In the following the arrangement and number 
of dots is arbitrary, but for simpler notations we consider 
only  dots with two electrons each. The bare confinement  potential 
 can vary from dot to dot.
If we introduce for each dot a  
c.m. coordinate ${\bf R}_n={\bf R}_n^0+{\bf U}_n$ with 
${\bf U}_n=(1/2) ({\bf u}_{n1}+{\bf u}_{n2}) $ and a relative coordinate 
${\bf r}_n= {\bf r}_{n2}-{\bf r}_{n1}= {\bf u}_{n2}-{\bf u}_{n1}$, 
then the total Hamiltonian decouples 
\begin{equation}
H=H_{cm}\left(\{ {\bf R}_n \}  \right)+
 \sum_n\; H_{rel,n}\left( {\bf r}_n \right)
\end{equation}
into a collective Hamiltonian 
\begin{equation}
H_{cm}=\frac{1}{2}
\Bigg \{ \sum_n
\frac{1}{2}
\left[
{\bf P}_n+
{2\over c} {\bf A} ({\bf U}_n)
\right]^2  
  + 2 \sum_{n,n'}\; {\bf U}_n \cdot
{\bf C}_{n,n'}
\cdot {\bf U}_{n'}
\Bigg \}
\label{H-array-cm}
\end{equation}
and a sum of individual intradot  Hamiltonians
\begin{equation}
H_{rel,n}=2 \left\{
\frac{1}{2}
\left[
{\bf p}+
{1\over 2 c} {\bf A}({\bf r}) \right]^2 +
\frac{1}{2}\; {\bf r} \cdot {\bf D}_n \cdot {\bf r}
+\frac{1}{2\;r}
\right \}
\label{H-array-rel}
\end{equation}
The force constant tensor ${\bf C}_{n,n'}$ of the collective Hamiltonian 
and the effective confinement tensor ${\bf D}_n$ of the decoupled intradot 
Hamiltonians contain both the bare confinement potential and 
a contribution from the dipole tensor from the interdot interaction 
\cite{Taut-QD-lattice}. 

\begin{figure}[htbp]
\centering
\includegraphics[width=12cm]{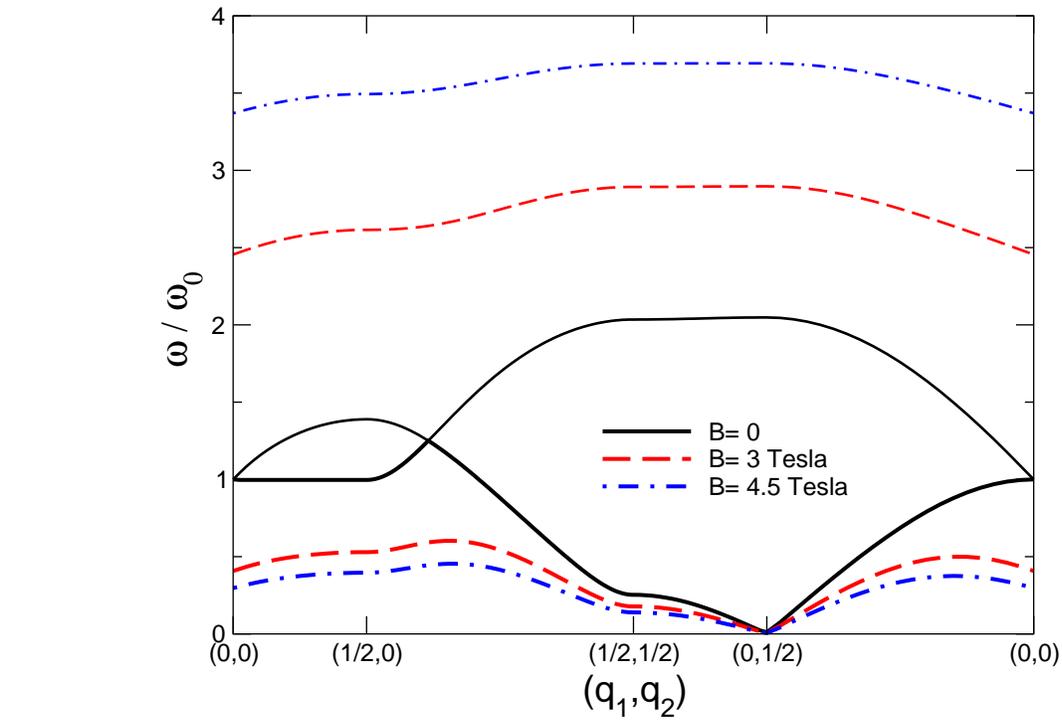}

\caption { Magneto-phonon dispersion in a rectangular quantum dot lattice 
with lattice constants $a_1/a_2=2$  for the critical interaction strength 
and for the three magnetic fields given in the legend.
The abszissa gives the magneto-phonon frequency $\omega$ in units of the 
bare confinement frequency $\omega_0$ of the dots. 
The wave vector $(q_1,q_2)$ shown 
on the ordinate varies along the edge of the irreducible Brillouin zone. 
}
\label{fig-magneto-phonon}
\end{figure}

The spectrum of the {\em intradot excitations} from (\ref{H-array-rel}) 
can be obtained with the methods for single quantum dots and its 
general features have been described in the previous sections. 
The Hamiltionian (\ref{H-array-cm}) describes {\em magneto-phonon excitations}.
If the strength of the interdot interaction reaches a critical value, 
a magneto-phonon mode can become soft indicating a lattice instability.
Such a case is shown in 
Fig.\ref{fig-magneto-phonon} for a rectangular pariodic lattice.
Solutions for a selection of dot dimers and periodic lattices are
given and discussed in some detail in Ref.\cite{Taut-QD-lattice}.

{\bf Acknowledgement}
This work was supported by the German Research Foundation (DFG)
in the Priority Program SPP 1145.

\end{document}